\documentclass[useAMS,usenatbib,useasmath]{mn2e}
\usepackage{epsfig}
\usepackage{longtable}
\usepackage{times}
\usepackage{amsmath}
\usepackage[usenames,dvipsnames]{color}

\newcommand{\Mdot}{\hbox{$\dot M$}}

\newcommand{\Rstar}{\hbox{$R_*$}}
\newcommand{\Rref}{\hbox{$R_{\hbox{\scriptsize \rm ref}}$}}
\newcommand{\Rmax}{\hbox{$R_{\hbox{\scriptsize \rm max}}$}}

\newcommand{\Msun}{\hbox{$M_\odot$}}
 \newcommand{\Msunyr}{\hbox{$M_\odot\,$yr$^{-1}$}}

\newcommand{\Teff}{\hbox{$T_{\hbox{\small eff}}$}}

\newcommand{\Vrot}{\hbox{$V_{\hbox{\scriptsize rot}}$}}
\newcommand{\Vcrit}{\hbox{$V_{\hbox{\scriptsize crit}}$}}
\newcommand{\tauross}{\hbox{$\tau_{\hbox{\scriptsize Ross}}$}}

\newcommand{\kms}{\hbox{km$\,$s$^{-1}$}}

\def\lesssim{\mathrel{\hbox{\rlap{\hbox{\lower4pt\hbox{$\sim$}}}\hbox{$<$}}}}
\def\gtrsim{\mathrel{\hbox{\rlap{\hbox{\lower4pt\hbox{$\sim$}}}\hbox{$>$}}}}

\newcommand{\loggc}{\hbox{$\log g_{\rm c}$}}
\newcommand{\vcl}{\hbox{$v_{\rm cl}$}}
\newcommand{\finf}{\hbox{$f_{\infty}$}}
\newcommand{\logL}{\hbox{$\log L/L_{\odot}$}}
\newcommand{\zpup}{\hbox{$\zeta$}~Pup}

\newcommand{\civ}{\mbox{C~{\sc iv}}}

\newcommand{\heii}{\mbox{He~{\sc ii}}}
\newcommand{\oiii}{\mbox{O~{\sc iii}}}

\newcommand{\niii}{\mbox{N~{\sc iii}}}
\newcommand{\niv}{\mbox{N~{\sc iv}}}

\newcommand{\lb}{$\lambda$}

\title[Influence of rotation on optical emission profiles of O stars]
{The influence of rotation on optical emission profiles of O stars}

\author[ D. John Hillier,  Jean-Claude Bouret, Thierry Lanz, Joseph R. Busche]
{D. John Hillier,$^1$\thanks{E-mail: hillier@pitt.edu} Jean-Claude Bouret, $^{2,3}$ Thierry Lanz,$^4$ Joseph R. Busche$^5$ \\
$^1$ Department of Physics and Astronomy \& Pittsburgh Particle physics, Astrophysics, and Cosmology Center (PITT PACC), University of Pittsburgh,  \\ Pittsburgh, PA 15260, USA \\
$^2$ Laboratoire d'Astrophysique de Marseille, CNRS-Universit\'e de Provence,
              P\^ole de l'\'Etoile Site de Ch\^ateau-Gombert, 38, rue Fr\'ed\'eric Joliot-Curie \\
13388 Marseille cedex 13, France\\
 $^3$ NASA/Goddard Space Flight Center, Greenbelt, MD 20771, USA \\
 $^4$ Laboratoire J.-L. Lagrange, UMR 7293, Universit\'e de Nice-Sophia Antipolis,
  CNRS, Observatoire de la Cote d'Azur, Boulevard de l'Observatoire, \\
  B.P. 4229, F-06304 Nice cedex 4, France \\
  $^5$ Wheeling Jesuit University, Wheeling, WV 26003 }

\begin{document}

\date{Accepted . Received }

\pagerange{\pageref{firstpage}--\pageref{lastpage}} \pubyear{2011}

\maketitle

\label{firstpage}

\begin{abstract}
We study the formation of photospheric emission lines in O stars and show
that the rectangular profiles, sometimes double peaked, that are observed
for some stars are a direct consequence of rotation, and it is unnecessary to invoke an enhanced density
structure in the equatorial regions. Emission lines, such as \niv\ \lb 4058
and the \niii\ \lb\lb4634-4640-4642 multiplet, exhibit non-standard ``limb darkening'' laws.
The lines can be in absorption for rays striking the center of the star and in
emission for rays near the limb. Weak features in the flux spectrum do not necessarily
indicate an intrinsically weak feature -- instead the feature can be weak because of
cancellation between absorption in ``core'' rays and emission from rays near the limb.
Rotation also modifies line profiles of wind diagnostics such as \heii\ \lb4686 and H$\alpha$
and should not be neglected when inferring the actual stratification, level and nature of wind 
structures.
\end{abstract}


\section{Introduction}

Many O stars exhibit rapid rotation that broadens their line profiles. In the absence of a wind, the influence of rotation can be accurately computed by integrating the intensities from the stellar disk with allowance for the projected velocity in the direction of the observer. For convenience, and speed, it is customary to allow for the effects of rotation by convolving the flux spectrum with a rotational broadening function  that accounts for limb darkening. This procedure, which assumes that the line profile does not change across the disk (i.e., the continuum and line have the same limb darkening laws) \citep[e.g.,][]{Gray92_book}, generally works very well. As our own tests have shown, the simple convolution procedure is sufficiently accurate for the majority of photospheric absorption profiles in O stars, although for the most precise work the disk integration procedure should be used.

In general, O stars exhibit a wind which modifies the photospheric  H$\alpha$ profile (possibly driving H$\alpha$ into emission) and generates numerous UV P~Cygni profiles. Because the wind is extended, and because the rotation rate declines with radius, the simple convolution technique is not valid. In such a case it is necessary to take the 2D structure of the wind into account. In the simplest case one can assume that rotation simply affects the velocity structure of the wind while maintaining an almost spherical structure, while for rapid rotation it is expected that the density structure will also be
altered. How the 2D density structure is altered is very complex, and depends on the rotation rate relative to the critical rotation rate, the rotation rate relative to the terminal wind velocity, and the closeness of the star to the Eddington limit. Depending on their values, it is possible to get either density enhancements in the equatorial or polar flows \citep{MM00_Edd_lim}.  A prolate wind structure is expected unless the number of lines (or more correctly the flux [line] mean opacity) increases sufficiently rapidly towards the equator to overcome the decreasing equatorial flux arising from gravity darkening \cite[e.g.,][]{MM00_Edd_lim}.
 
In many O stars intense \niii\  \lb4634--4642 emission is seen and is used as a criterion to
define the so-called ``f" class \citep{Wal71_Of, SMW11_class}. When resolved, these lines can show a rectangular (or trapezoidal-like) profile\footnote{For simplicity we refer to these profiles as non-Gaussian, while other profiles will be referred to as Gaussian, even if they do not exhibit extended wings.}, with the stronger components exhibiting a  double peaked structure. In the spectra of O supergiants presented by \cite{BHL12_Osg}, \zpup, HD~16691, and HD~210839 show broad, non-gaussian profiles for the \niii\ multiplet, although HD~15570, HD~163758 and HD~192639 exhibit more gaussian-like profiles.  The mechanisms driving the \niii\ lines into emission have been discussed previously by \cite{RPN11_NIII}. The dominant mechanism depends on the precise stellar parameters -- both continuum fluorescence (e.g., the Swings mechanism, \citealt{BM71_NIII}) and dielectronic recombination (previously discussed for these lines by \citealt{MHC72_NIII}) can be important. \cite{RPN11_NIII} also indicate that the strength of the lines can be influenced by interactions between the \niii\ and \oiii\ resonance lines.

In this paper we discuss the profiles of photospheric emission lines in O stars, and show that
rotation can produce rectangular-like profiles that often exhibit double-peaked profiles. The
phenomena is not restricted to the \niii\  \lb4634--4642 complex --- in  \zpup \niv\ \lb 4058 and  Si\,{\sc iv} \lb\lb 4089,4116 also exhibit non-gaussian profiles. We show that the observed line profiles are a direct consequence of the lines being photospheric and being in emission. The lines do not show  ``classic'' limb darkening -- rather the intensity distribution across the star can be flat, and may even show limb brightening. In addition, we further examine the influence of rotation on the H$\alpha$ and He\,{\sc ii} 4686 line profiles. Rotation produces observable effects on the line profiles, and these effects are seen. Because of the influence of rotation on the wind velocity law (which is now 2D) wind line profiles potentially depend on both $v$ and $\sin i$ rather than $v \sin i$. To model these emission lines on O stars, it is absolutely essential to properly account for the effects of rotation in O stars that exhibit rapid rotation. It is especially true for rapid rotators  such as $\zeta$ Pup which has $v \sin i \sim  210 \,\kms$.

\section{Previous Work}

The influence of rotation on O star profiles has been the subject of several different studies.
\cite{PP96_2D} studied the influence of rotation on H$\alpha$. Their work was mainly
concerned with the influence of rotation on the wind profile. Rotation directly influences
the velocity structure (the velocity along a given sight line is not necessarily 
monotonic) and can potentially alter the density structure of the wind through 
direct means (as in the wind-compressed disk (WCD) model of \cite{BC93_WCD}) or
indirectly by introducing a latitudinal variation of temperature and surface gravity
which then causes a latitudinal variation of the mass loss rate and the velocity law. 

In the work of \cite{PP96_2D} emission in H$\alpha$ was significantly modified by the enhancement of density in the equatorial plane arising from the WCD effect. However, this effect was likely substantially overestimated --- \cite{OCG98_WCD_inhib} showed that non-radial forces will inhibit both the formation of a wind-compressed disk and a wind-compressed region. As a consequence, the most important influence of rotation on the wind arises from rotation introducing a latitudinal dependence of temperature and surface gravity, but this effect remains relatively small unless $\Vrot > 0.8 \Vcrit$. Two  potentially important effects for wind lines are the decrease of $V_\phi$ with $r$ which
reduces the influence of rotational broadening on wind lines, and the resonance-zone effect which arises because rotation introduces a non-monotocity in the velocity law along some sightlines \citep{PP96_2D}. Given the current uncertainties in our understanding of 1D winds, it is fair to say that the structure of 2D winds, and the azimuthal variation of density and mass-loss, is still very uncertain.

Another study of the influence of rotation on line profiles in O stars was made by
\cite{BH05_2D}. They examined a range of profiles, showed that 
the absorption profiles were adequately modeled via the convolution technique,
and confirmed the important influence of rotation on H$\alpha$ and \heii\ \lb 4686
line profiles. While the analysis was general, and applied to the whole spectrum,
they did not study the formation of photospheric emission lines.

\section{Technique}

To study the influence of rotation we use the 2D code developed by \cite{BH05_2D}. This
code computes the formal solution (and hence line profiles) for axisymmetric 2D winds.
It uses as input 1D opacities and emissivities computed with the 1D formal solution code,
CMF\_FLUX \citep{BH05_2D}. It has been parallelized, and runs efficiently on workstations
with multiple processors.

In the present calculation we assume that rotation does not influence the density structure
--- this assumption is a reasonable first approximation since the rotation rates are less than
50\% of the critical rotation rate, and much less than the wind terminal velocities. We also neglect the latitudinal variation of temperature and surface gravity with latitude --- these are likely not to have a large effect since the rotational velocity of most O stars is generally  a factor of 2 (or more) less than the critical rotational velocity. Following the work of \cite{BH05_2D} we assume that the star rotates as a solid body below  20\,\kms, and that angular momentum about the center of the star is ``conserved"  at higher velocities (that is, $V_\phi \propto R(V=20\,\kms)/r$).  Following the work of \cite{OCG98_WCD_inhib} we assume non-radial forces inhibit disk formation and hence set the polar velocity component to zero. The last two assumptions only influence the formation of wind lines -- they do not affect photospheric lines. For simplicity,
we neglect the influence of rotation on the non-LTE populations -- such effects are likely weak, and have been discussed by \cite{PP00_2D}.

\section{Observations and model}

To illustrate the influence of rotation we use $\zeta$~Puppis which is a fast rotator with 
$v  \sin i \sim 210\kms$. This star has been extensively analyzed, most recently by \cite{BHL12_Osg}.
\cite{NHP11_Lband} have also extensively analyzed this star and obtained similar results.
The model parameters we adopt are from the paper of \cite{BHL12_Osg}, and are as follows:
$\Teff=40\,000$\,K, $\loggc =3.64$ (centrifugally corrected value of $\log g$), $\Mdot=2.0 \times10^{-6}$\,\Msunyr, $\logL=5.91$, $\finf=0.05$, $\vcl=100\,$\kms, N(He)/N(H)\,$=0.16$, Z(C)\,$=2.86 \times 10^{-4}$, Z(N)\,$=1.05 \times 10^{-2}$ and Z(O)\,$=1.30 \times 10^{-3}$ (Z denotes mass fraction).\footnote{Expressed as number fractions: N(C)/N(H)$=4.0 \times 10^{-5}$, N(N)/N(H)$=1.2 \times 10^{-3}$ and
N(O)/N(H)$=1.4 \times 10^{-4}$.} $\finf$ and $\vcl$ control the clumping in the wind and have little influence on the photospheric profiles. For illustration purposes, we use the same observational data (which covers 3800 to 8800\,\AA\, at a resolution of 48,000 and a signal-to-noise in excess of 100 to 1 per pixel)  as presented in that paper.

$\zeta$~Puppis is an ideal candidate for the study. Its parameters are well known (except for uncertainties induced by the the uncertainty in its distance; see discussion in \citealt{SR08_Ostars}), and from variability studies it is most likely observed edge on (i.e., rotation axis perpendicular to line of sight; \cite{HPM95_zeta_pup}). It exhibits several emission lines such as \niv\ \lb 4058, \niii\ \lb 4634-4642, and Si\,{\sc iv} \lb\lb 4089,4116 that show non-gaussian profiles - generally rectangular but possibly exhibiting enhanced emission near the extremes of the emission profile. Further, \civ\ 5801, 5812 exhibits complex profiles while both H$\alpha$ and He\,{\sc ii} are in emission. 

\begin{figure*}
\includegraphics[scale=0.45, angle=-90]{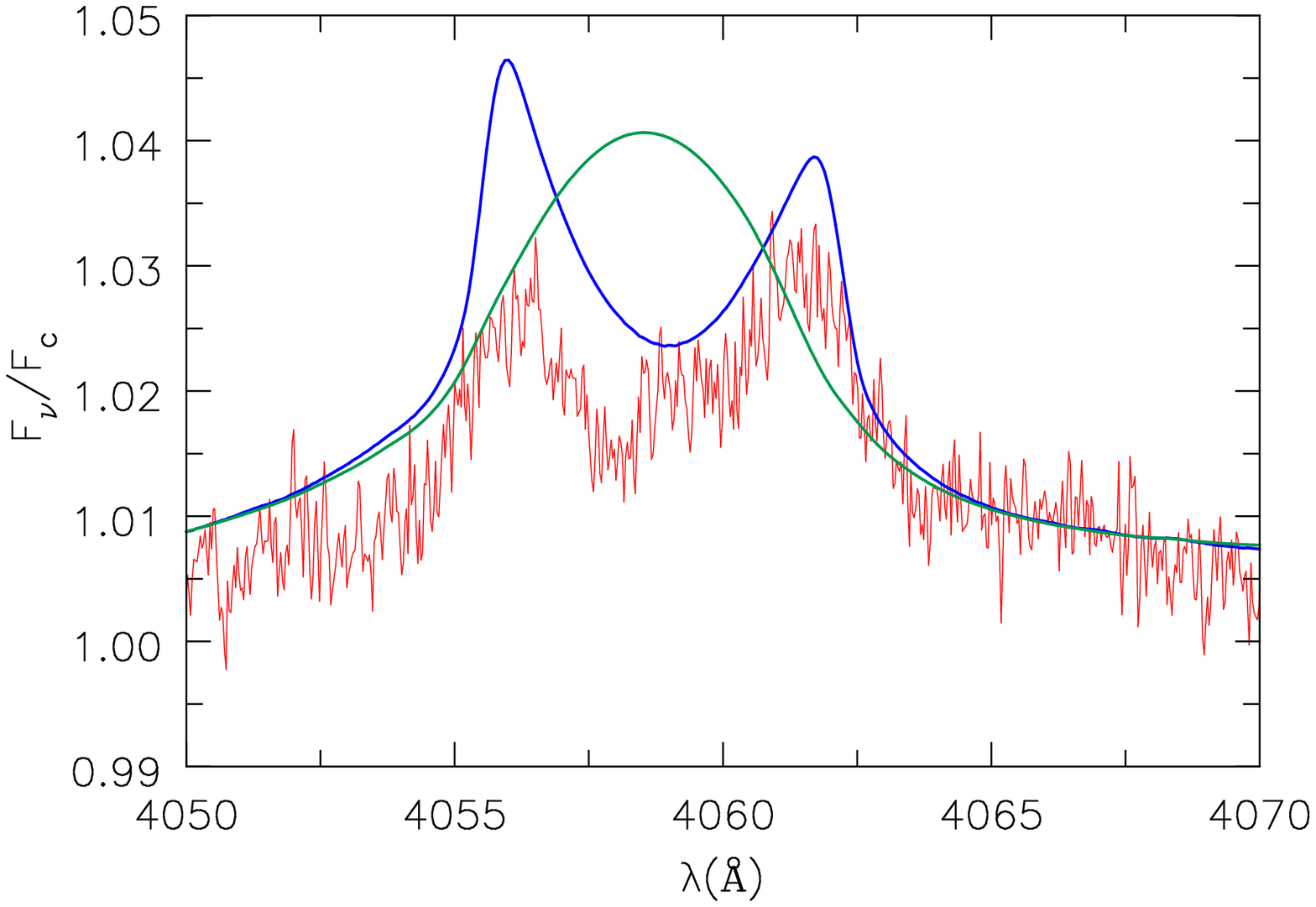}
\hspace{0.8cm}
\includegraphics[scale=0.45, angle=-90]{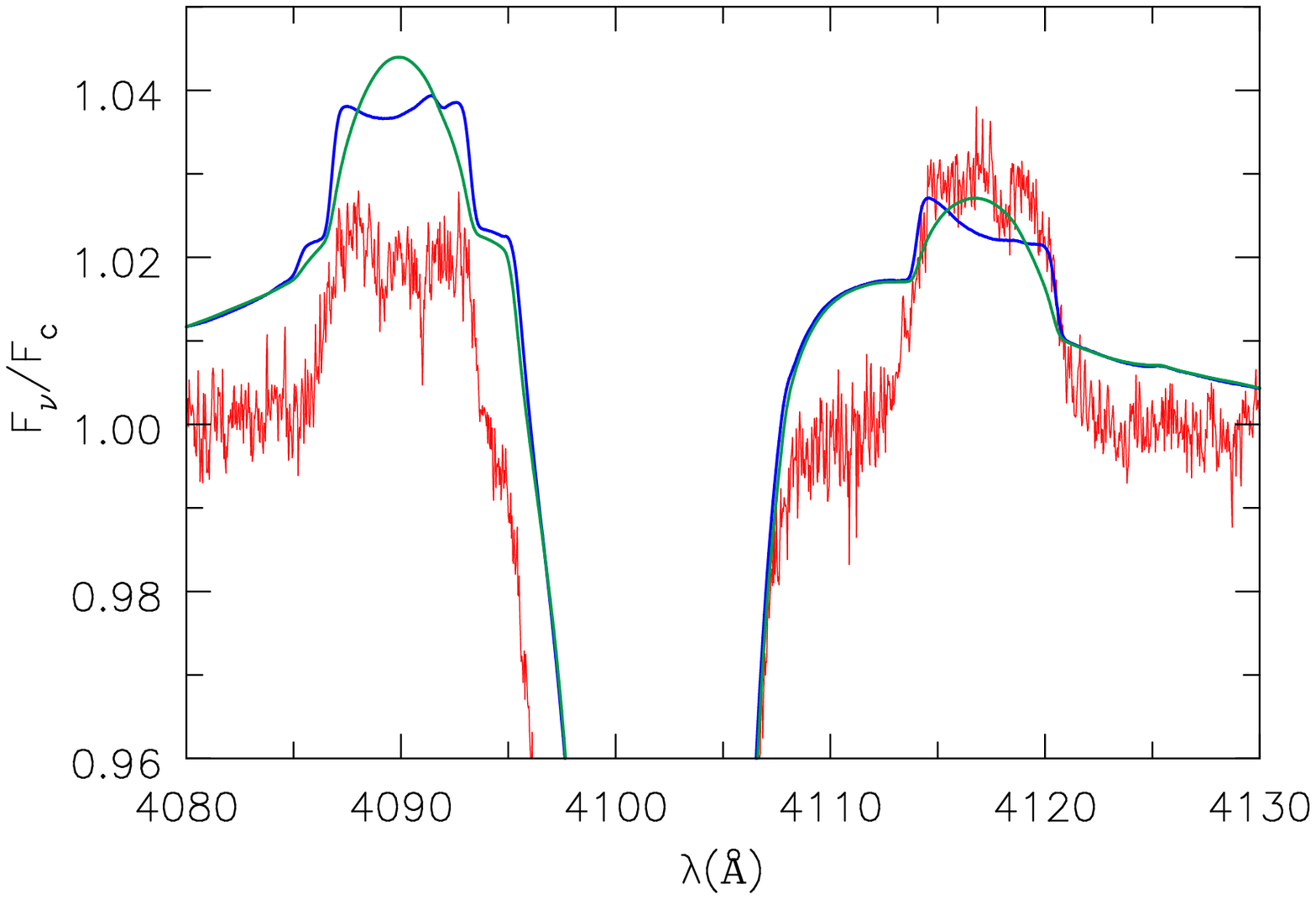} \\
\vspace{0.3cm}
\includegraphics[scale=0.45, angle=-90]{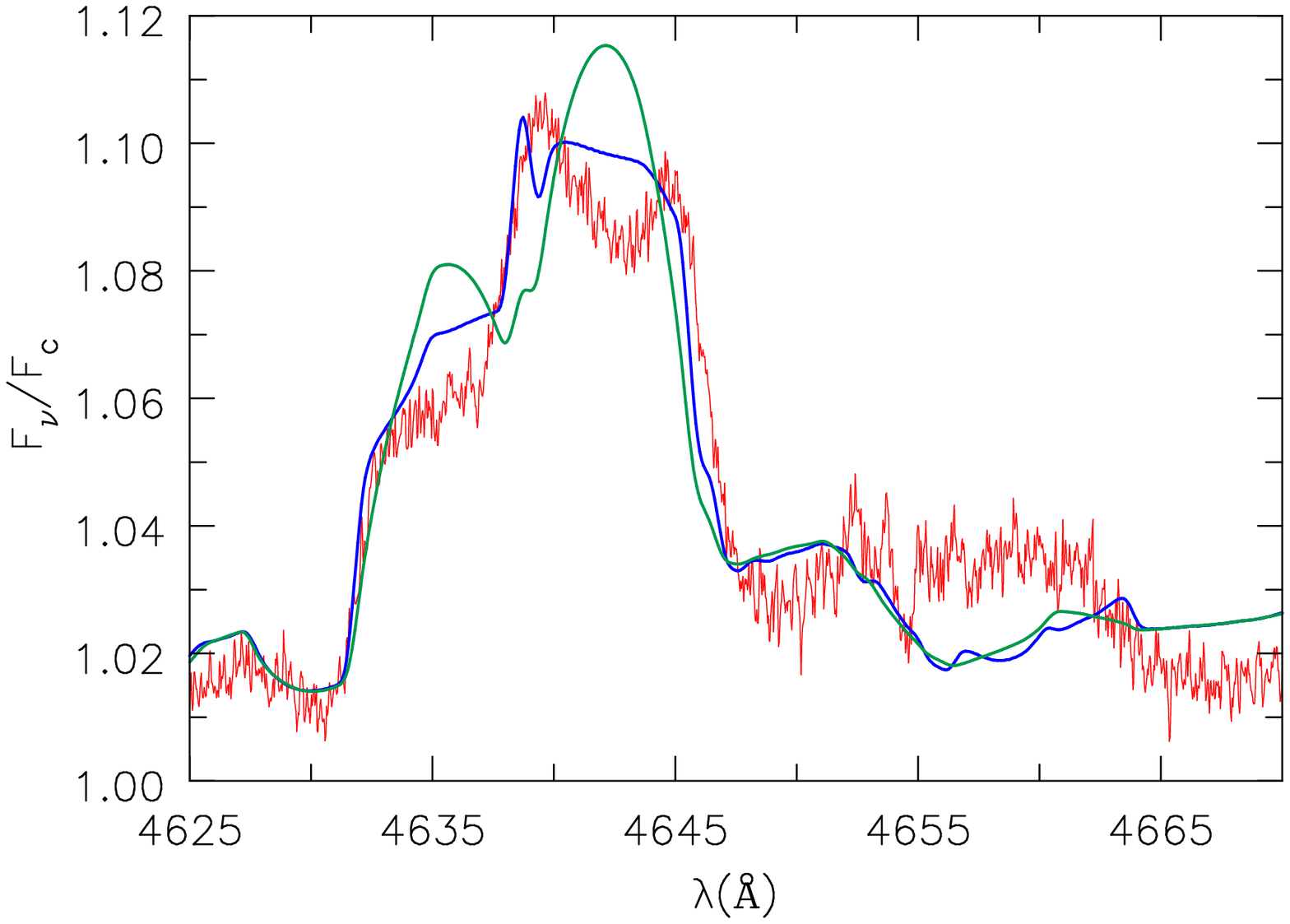}
\hspace{1.1cm}
\includegraphics[scale=0.45, angle=-90]{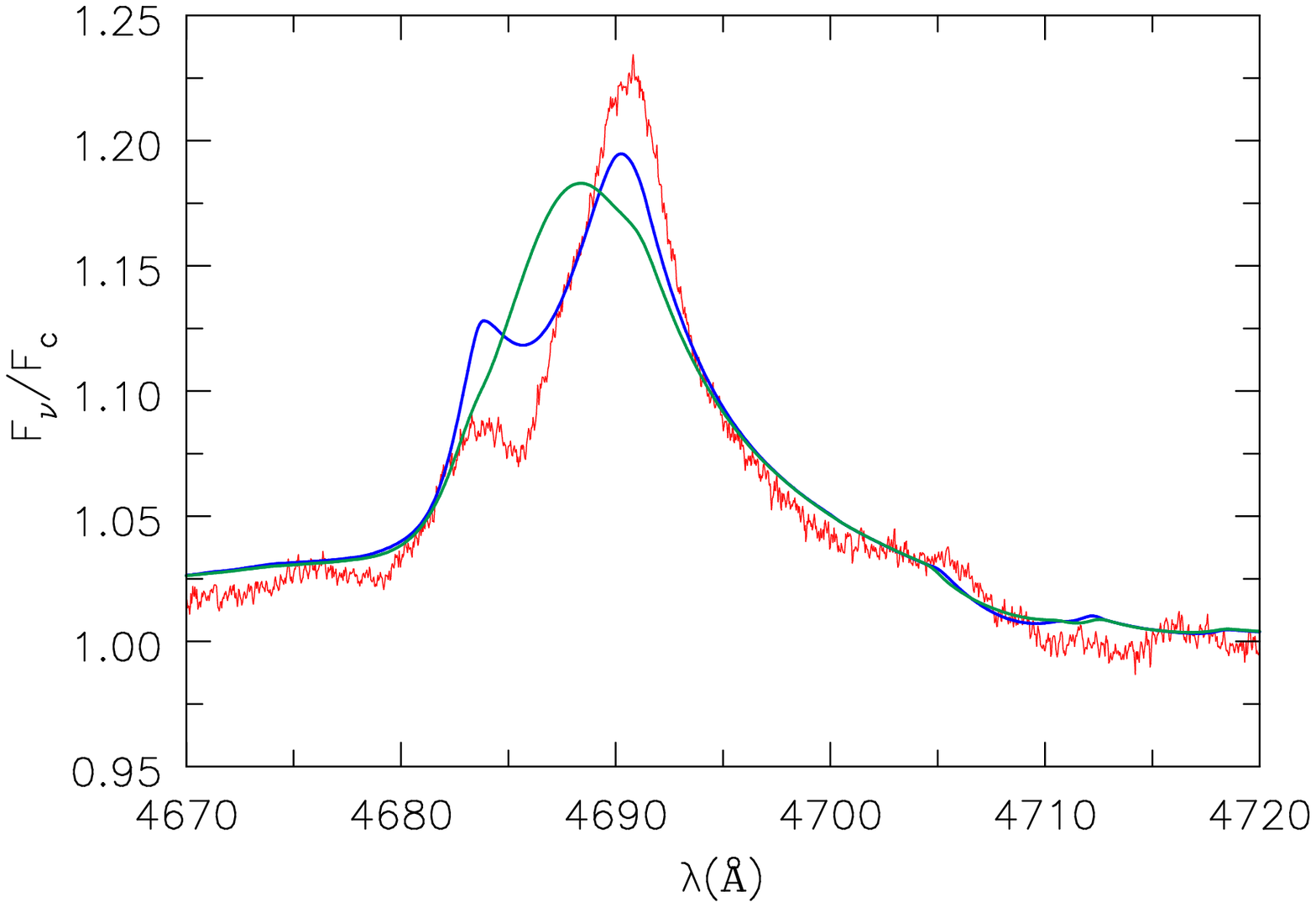} \\
\vspace{0.3cm}
\includegraphics[scale=0.45, angle=-90]{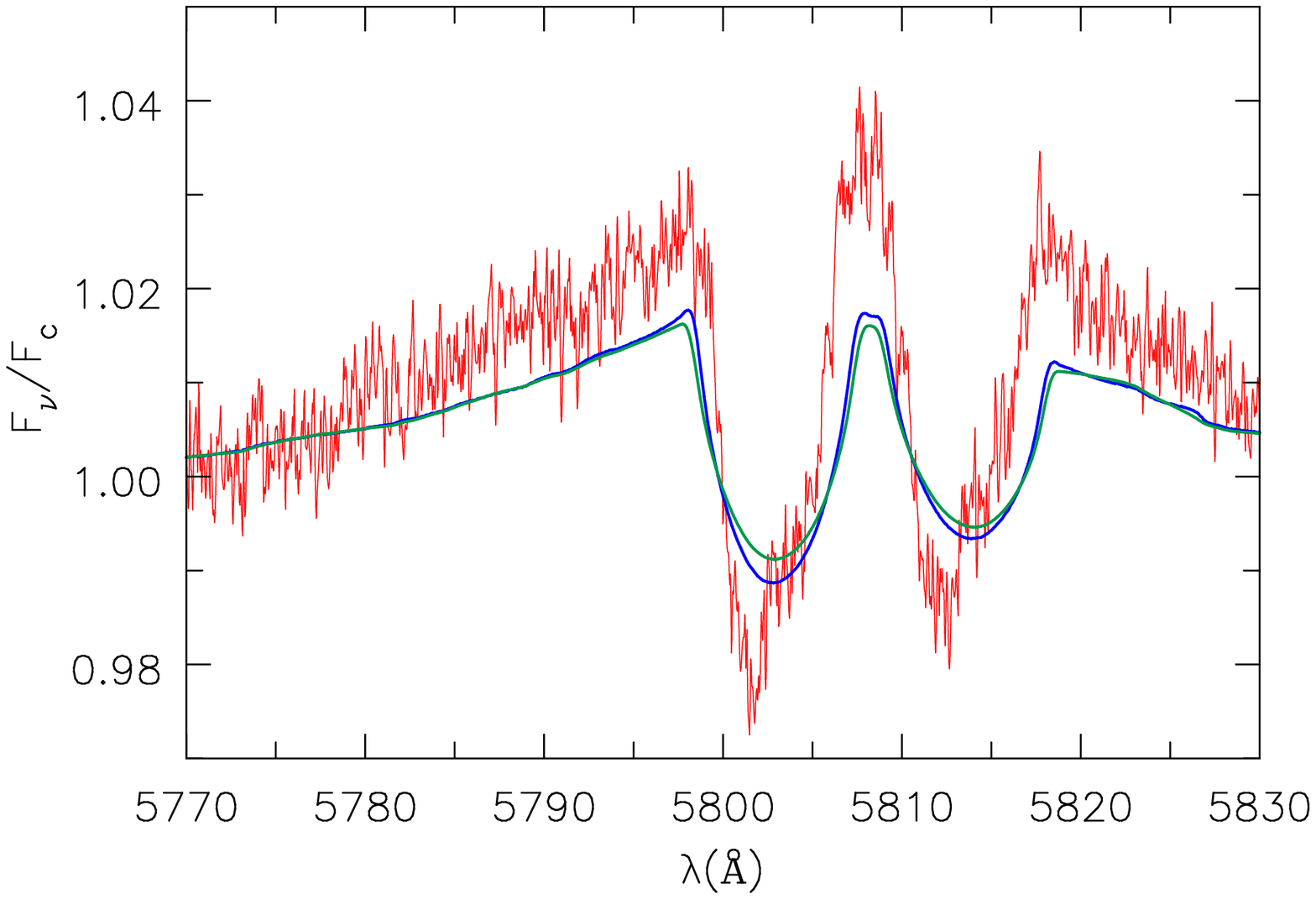}
\hspace{0.8cm}
\includegraphics[scale=0.45, angle=-90]{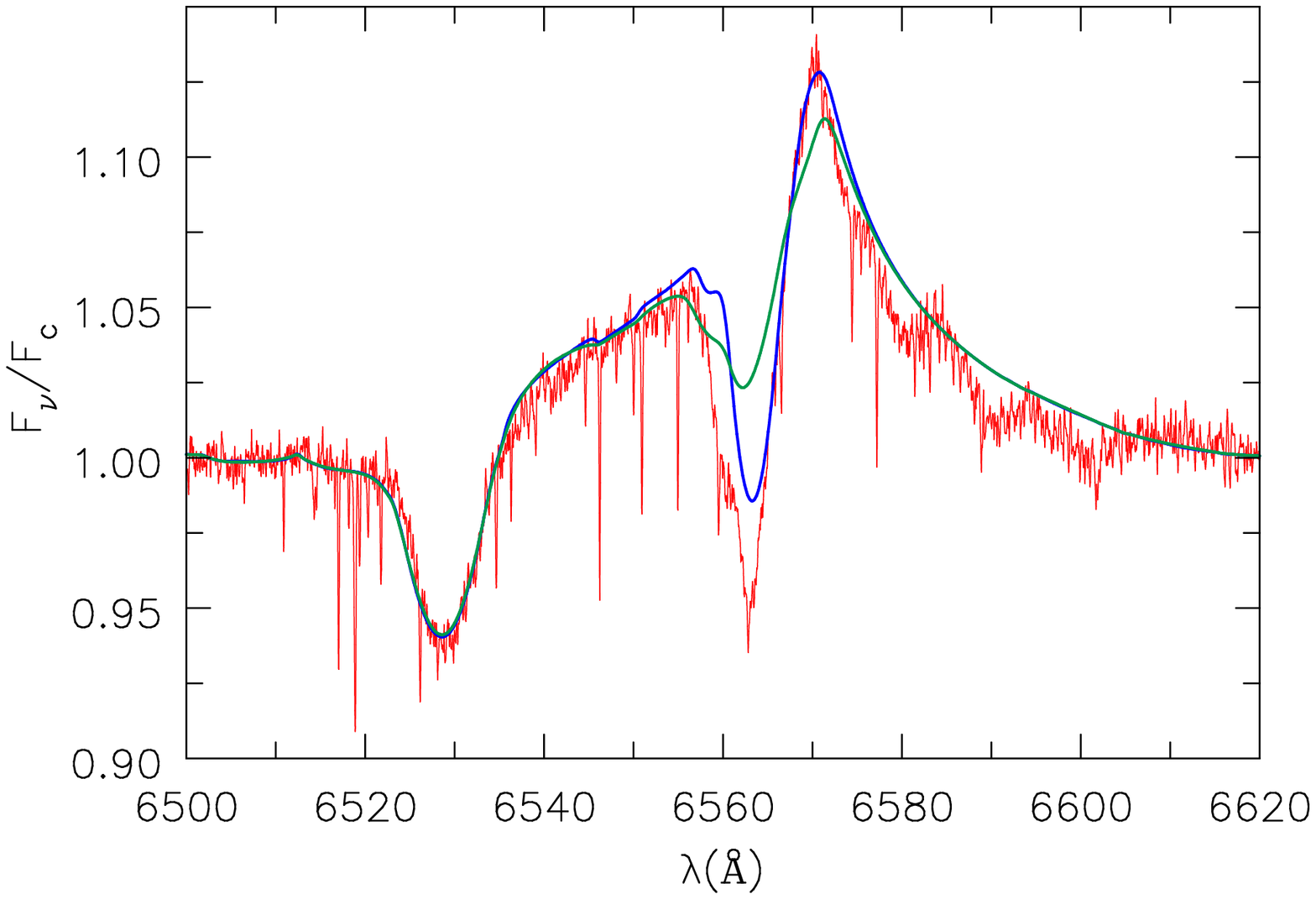}
\caption[]{Illustration of a sample of line profiles. The red curve shows the spectrum of $\zeta$~Pup, the blue curve shows the spectrum computed using the formal integral of the 2D model with $v \sin i=$\,240\,\kms, and the green curve show profiles obtained through convolution of the 1D-flux spectrum with $v \sin i=$\,240\,\kms. From left to right, top to bottom we show \niv\ \lb 4058, Si\,{\sc iv} \lb\lb 4089,4116, \niii\ \lb \lb 4634-4640-4642, \heii\ \lb 4686,  \civ\ \lb\lb 5081, 5812 and H$\alpha$.  In all cases the profiles computed with the 2D model show better agreement than the profiles obtained via convolution with that observed. In particular notice the distinct non-gaussian profiles associated with \niv\ \lb 4058, Si\,{\sc iv} \lb\ 4089, 4116, and \niii\ \lb \lb 4634-4640-4642. On this plot wavelengths are vacuum, but we use air wavelengths (as is the usual practice) when identifying lines.}
\label{fig_prof}
\end{figure*}

Observed profiles for a selection of lines in \zpup\ are shown in Fig,~\ref{fig_prof}. Also shown are the profiles obtained from a simple convolution of the 1D model flux with the rotation profile, and the profile computed from a full integration of the intensities across the disk of the star (hereafter disk profiles). For simplicity we adopted a macroturbulent velocity\footnote{The use of a (constant) macroturbulent velocity is, of course, another approximation which will modify line profiles and hence influence conclusions that can be drawn from them.} of 0\,\kms (\citealt{BHL12_Osg} adopted 90\,\kms) and a rotation velocity of 240\,\kms\ (and $\sin i=1$) to better match the width of the rectangular emission profiles rather than the 210\,\kms\ adopted by \cite{BHL12_Osg}. As readily apparent, the disk profiles generally show much better agreement with the observed profiles than do the profiles obtained through convolution. In particular, \niv\ \lb 4058 exhibits a double peaked profile, as observed. Similarly, the \niii\ lines show rectangular profiles. Lines affected by the wind also show improvements in profile shapes ---  He\,{\sc ii} 4686 exhibits evidence for ``absorption'' on the blue side which is clearly seen in the observation but is not present in the profiles created by convolution.
Profiles computed assuming a higher rotation velocity of $480\,\kms$ and $\sin i=0.5$ so that $v \sin i = 240$ are essentially identical to those computed using the lower rotation velocity. In reality, since the wind
and star properties would change with such a high  rotation rate, the profiles would not be identical.

It is not surprising that the convolution procedure does not work well for H$\alpha$ and He\,{\sc ii} $\lambda 4686$. Both lines  are significantly influenced by the wind, and line transfer in the wind will be significantly affected by the presence of rotation.

From Fig,~\ref{fig_prof} is is apparent that predicted disk profiles do not agree precisely with those observed. This likely arises because of errors in the adopted stellar parameters and abundances, errors in the atomic data, and missing physics. A full discussion of the  uncertainties associated with the parameters of \zpup\ has been given by \cite{BHL12_Osg}. As the purpose of this paper is to understand the physics giving rise to the observed rectangular profiles, and to re-examine the influence of rotation, we do not attempt to readjust model parameters. We note, for example, that the absorption associated with H$\alpha$ is sensitive to the radial variation of clumping with radius \citep{HLH03_AV83,PMS06_clump,NHP11_Lband} and porosity effects \citep{SPF11_clumps}. Our own work on O supergiants \citep{BHL12_Osg} also shows that it is model sensitive (it varies significantly when the stellar parameters are varied over their error range).

\begin{figure}
\centering
\includegraphics[scale=0.45, angle=-90]{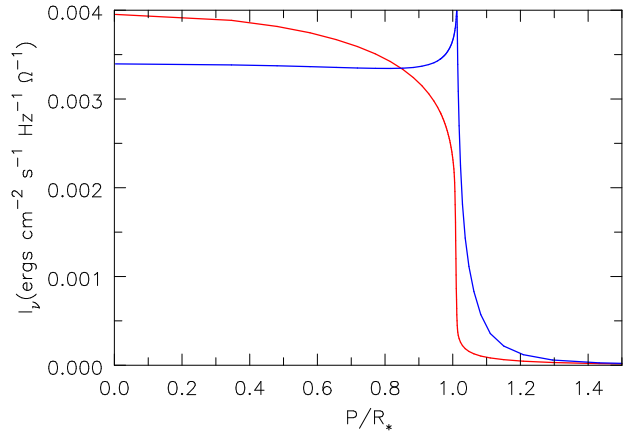}
\caption[]{Illustration of the ``limb darkening law" for the \niv\ \lb 4058 line at line center (blue curve) and the adjacent continuum (\lb 4056; red curve).
The continuum shows a classic limb-darkening law. Conversely, because of non-LTE effects, the \niv\ \lb 4058 line intensity is almost constant
across the star and brightens near the limb.}
\label{fig_niv_limb}
\end{figure}

To understand the observed emission line profile, and why convolution
does not work we consider \niv\ \lb4058. In Fig.~\ref{fig_niv_limb}  we show the intensity variation
at line center (no rotation) and in the adjacent continuum as a function of
impact parameter. As apparent, the intensity variation in the continuum and the line
is very different --- the continuum shows ``classic'' limb darkening, while the
intensity in \niv\ \lb4058 is almost flat. Since the ``limb darkening laws" for the continuum
and line differ substantially, we cannot use the simple convolution procedure to obtain the
profile when the star is rotating. Rather, the observed line profile must be obtained by
integrating the intensities across the disk.
 
\begin{figure*}
\centering
\includegraphics[scale=0.45, angle=-90]{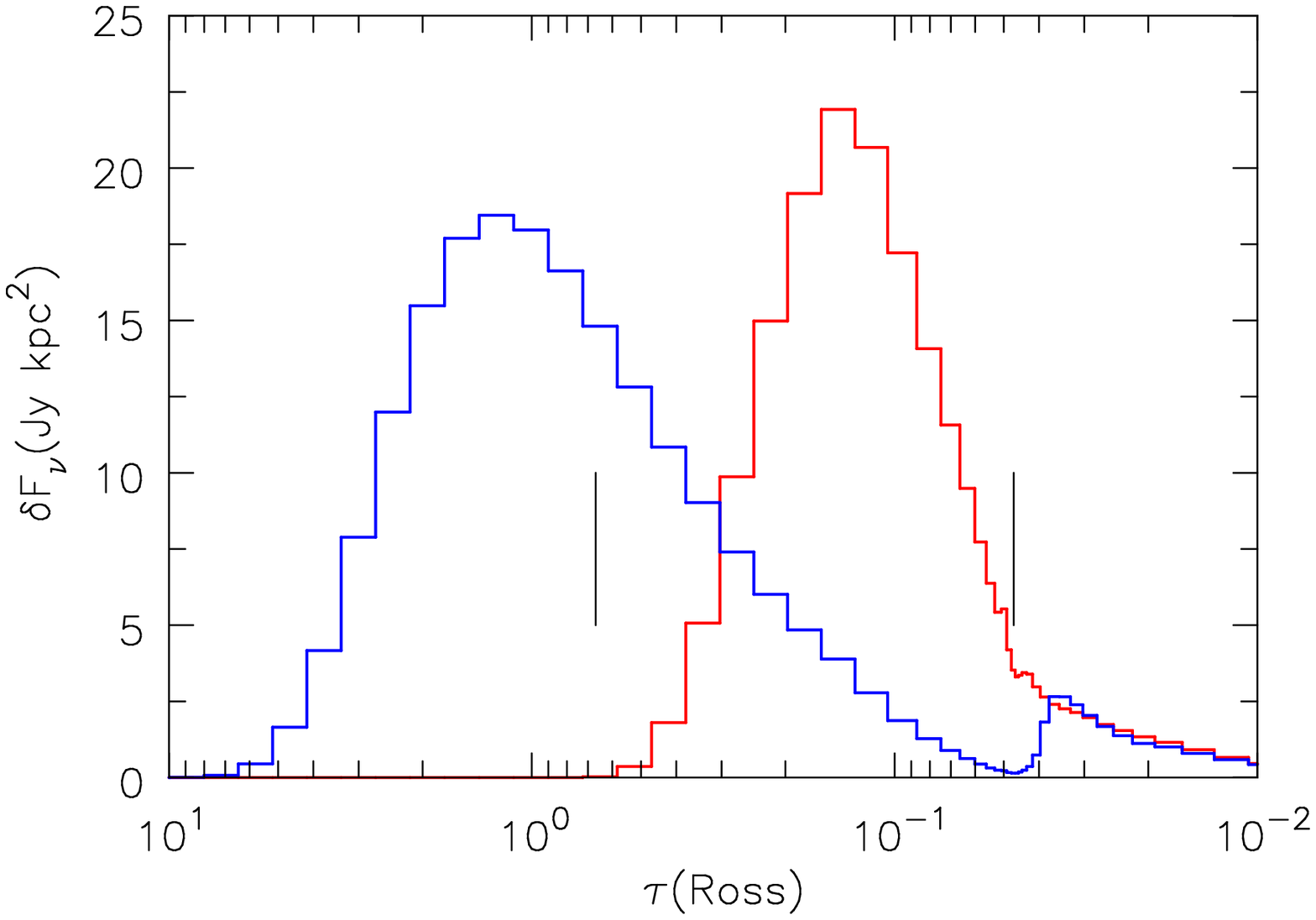}
\hspace{0.8cm}
\includegraphics[scale=0.45, angle=-90]{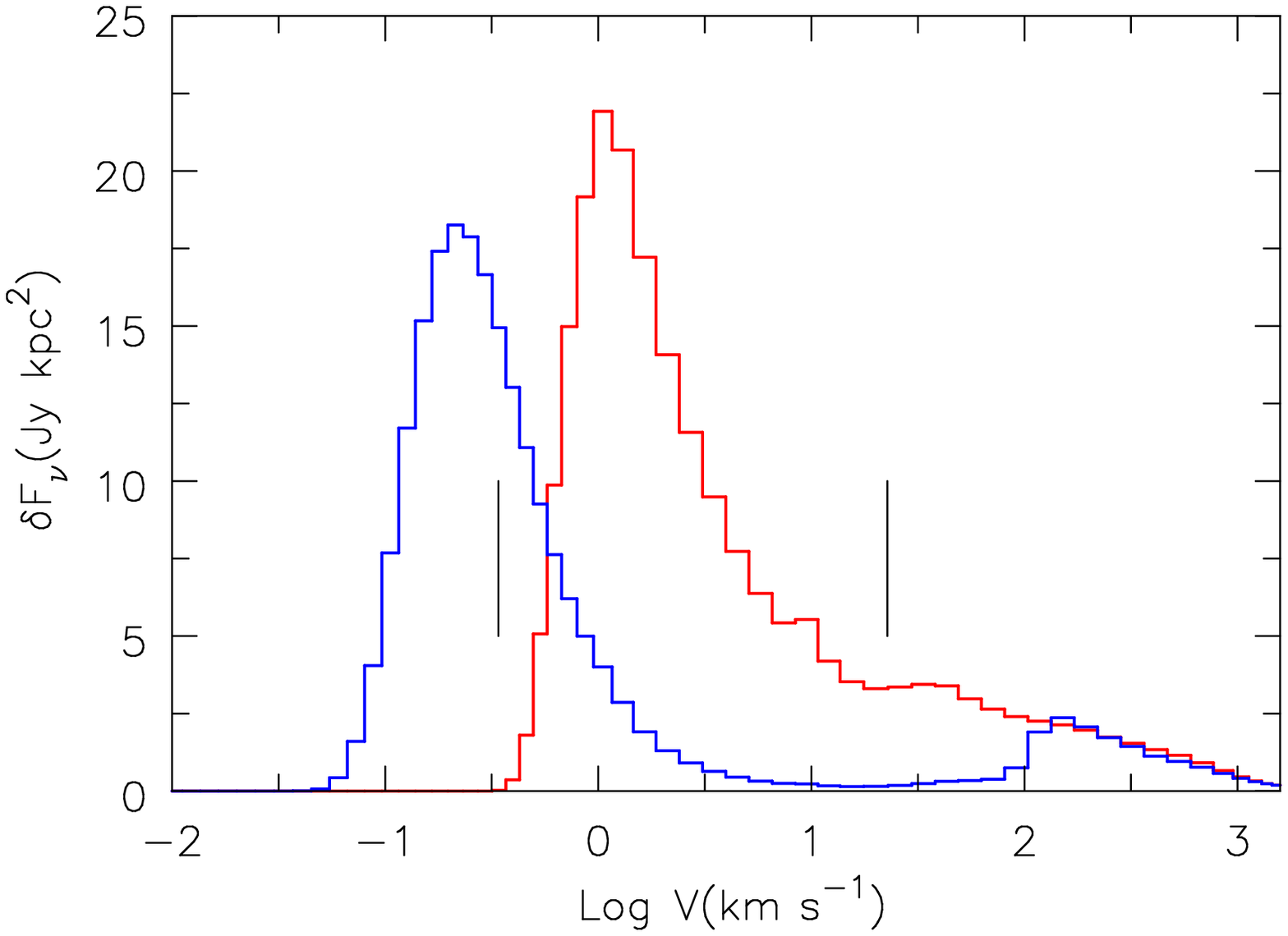}
{\caption[12cm]{Illustration of the origin of the \niv\ \lb 4058 line at line center (blue curve) and the adjacent continuum (\lb 4057; red curve) in $\tau$ and (radial) velocity space for a 1D non-rotating
star. Each data value (bin) provides the contribution to the total flux from that pixel -- the total flux is the sum over all pixels. The two vertical lines in each figure indicate the location of \tauross$=2/3$ (left line) and the location where the wind velocity matches the isothermal sound speed ($\sim 23$\,\kms).}
\label{fig_niv_orig}}
\end{figure*}

From Fig.~\ref{fig_niv_limb} it is also apparent that the wind makes relatively little contribution
to \niv\ \lb4058. \niv\ \lb 4058 is a photospheric emission line, and is in emission because of
non-LTE effects. In Fig.~\ref{fig_niv_orig} we illustrate the origin of the \niv\ \lb 4058 as a function
of the Rosseland mean optical depth, and velocity. This diagram also confirms that \niv\ \lb 4058
is photospheric --- the bulk of the line originates inside 10\,\kms.

\begin{figure*}
\begin{minipage}[t]{0.48 \linewidth}
\centering
\includegraphics[scale=0.42, angle=-90]{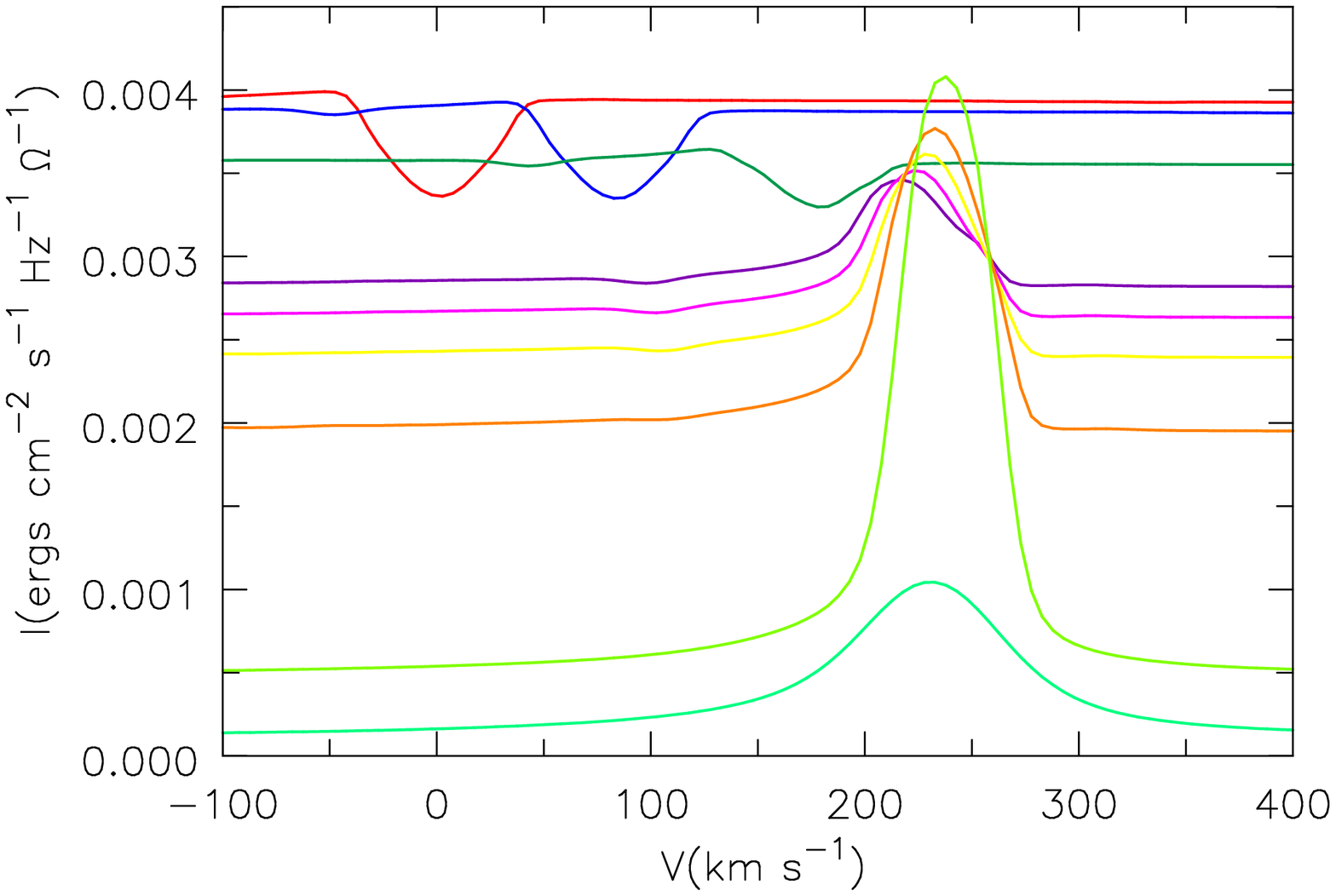}
\hspace{1.5in}\parbox{3.0in}{\caption{Intensity along the equator as a function of impact parameter for \niv\ \lb4058. The rotation velocity is 240\,\kms. From top to bottom we show the intensity for $p/\Rstar=$ 0.0, 0.039, 0.733, 0.965, 0.983, 0.997, 1.007, 1.0127, and 1.062 where \Rstar\ is the radius of the star at a Rosseland mean optical depth of 2/3. Notice how the line is in absorption near the center of the star but goes into emission near the limb. We only show the curves for half the star -- curves for the equatorial
region rotating towards the observer are similar (although not identical) to those shown above.}
\label{fig_niv_id}  } 
\end{minipage}
\begin{minipage}[t]{0.48 \linewidth}
\centering
\includegraphics[scale=0.42, angle=-90]{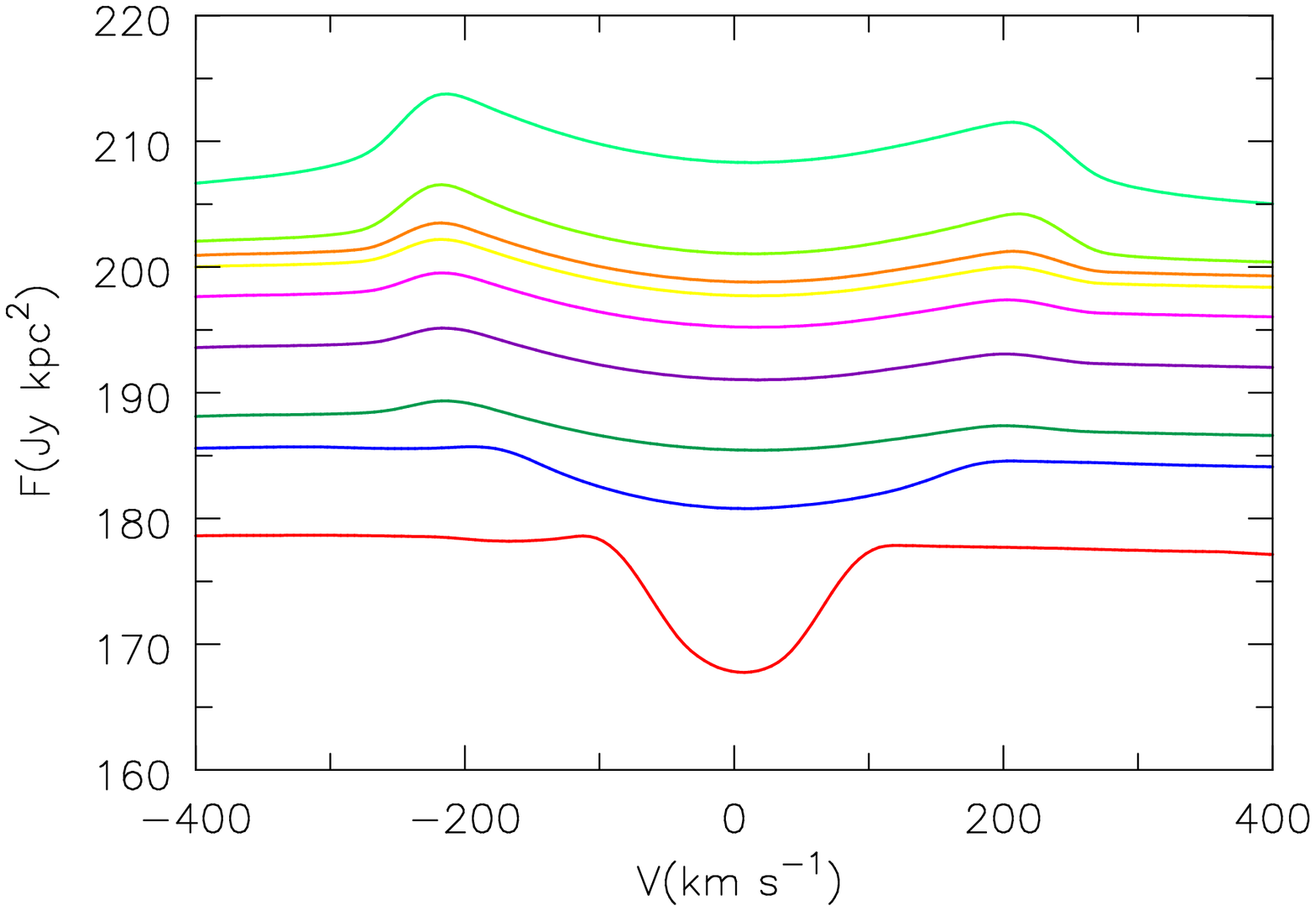}
\hspace{1.5in}\parbox{3.0in}{\caption{Plot showing the spectrum computed by integrating from
$p=0$ to $p(i)$ where $p(i)/\Rstar =$ 0.039, 0.733, 0.965, 0.983, 0.997, 1.007, 1.0127, 1.062 and 100 (=\Rmax/Rstar). The curves for $p(i)/\Rstar = 0.039$ (bottom) and $p(i)/\Rstar = 0.733$ have been scaled upwards by factors of 7.5 and 1.7, respectively. These plots, together with Fig.~\ref{fig_niv_id}, illustrate how the observed profile is a complicated blend of absorption and limb emission. }
\label{fig_niv_sp} }
\end{minipage}
\end{figure*}

To better understand why \niv\ \lb 4058 exhibits a non-gaussian profile we can examine the center-to-limb variation of the intensities as a function of velocity, and this is shown in Fig.~\ref{fig_niv_id} for a range of impact parameters. As readily apparent, the \niv\ \lb 4058 profile shows substantial variation with impact parameter -- for the core ray the line is in absorption while it is strongly in emission near the limb. Since the observed flux is a complicated weighting of the intensities, we illustrate in Fig.~\ref{fig_niv_sp} the growth in the spectrum with impact parameter. When integrated over the inner core rays (for all azimuthal angles) we see an absorption line spectrum, but as we extend the integration zone to higher impact parameters we begin to see emission, and eventually we get the double peaked profile, similar to what is observed.

\begin{figure*}
\begin{minipage}[t]{0.48 \linewidth} 
\centering
\includegraphics[scale=0.42, angle=-90]{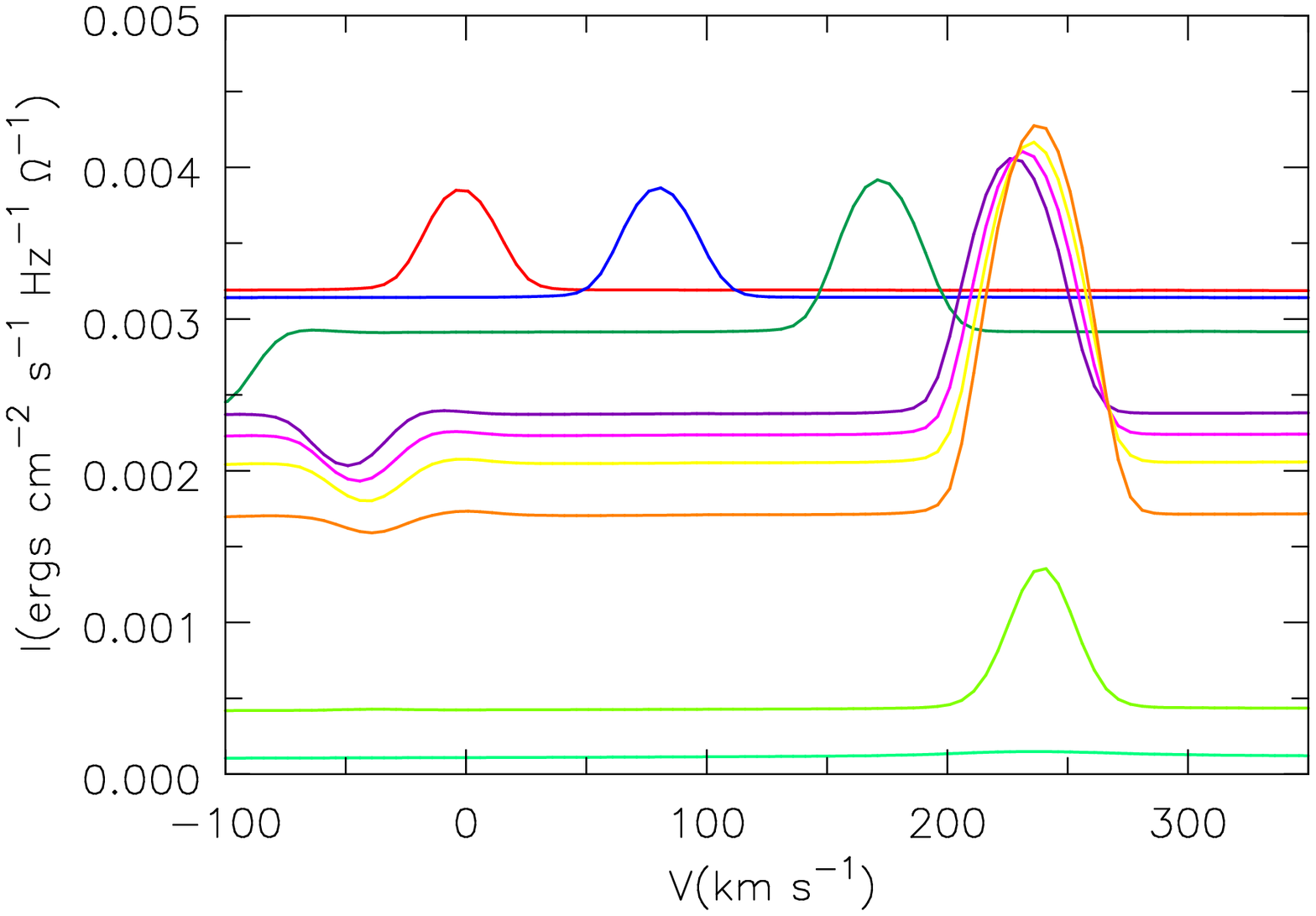}
\hspace{1.5in}\parbox{3.0in}{\caption{As for Fig.~\ref{fig_niv_id} but illustrating \niii\ \lb 4634.
The strength of the emission relative to the continuum is significantly enhanced towards the
limb of the star.}
\label{fig_niii_id} }
\end{minipage}
\begin{minipage}[t]{0.48 \linewidth}
\centering
\includegraphics[scale=0.42, angle=-90]{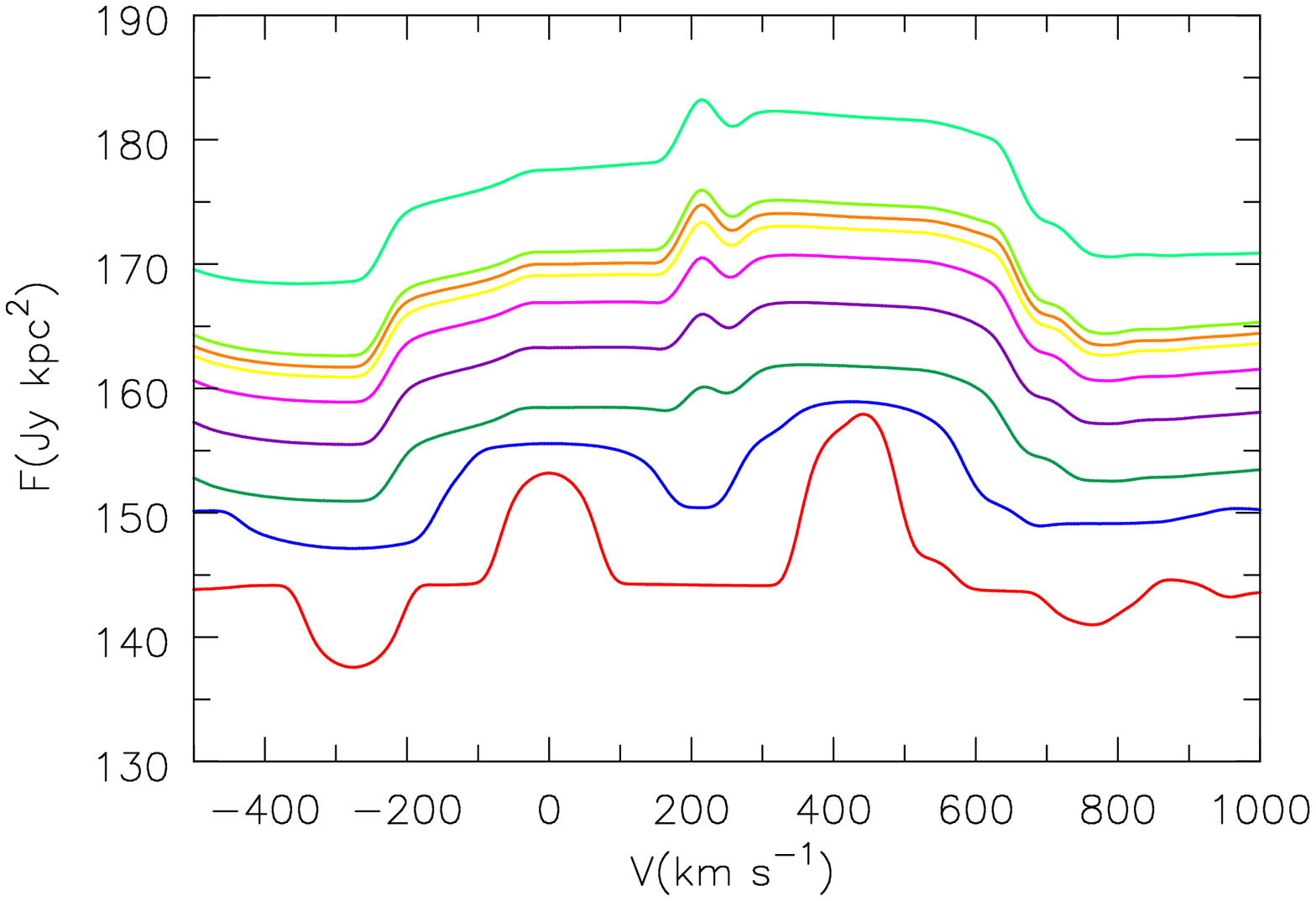}
\hspace{1.5in}\parbox{3.0in}{\caption{As for Fig.~\ref{fig_niv_sp} but illustrating  \niii\ \lb 4634.
At $\sim$400\,\kms\ we see  a blend of two other \niii\ lines (\lb\lb 4641, 4642) that belong to the same
multiplet.}
\label{fig_niii_sp} } 
\end{minipage}
\end{figure*}

Another line that exhibits a rectangular-shaped profile is \niii\ \lb 4634. While the line is in emission
at all impact parameters (Fig.~\ref{fig_niii_id}), it is much more in emission at the limb. It is this enhanced 
emission from ``large" impact parameters that changes the profile shape from Gaussian to
rectangular (Fig.~\ref{fig_niii_sp}).

\begin{figure*}
\begin{minipage}[t]{0.48 \linewidth} 
\centering
\includegraphics[scale=0.42, angle=-90]{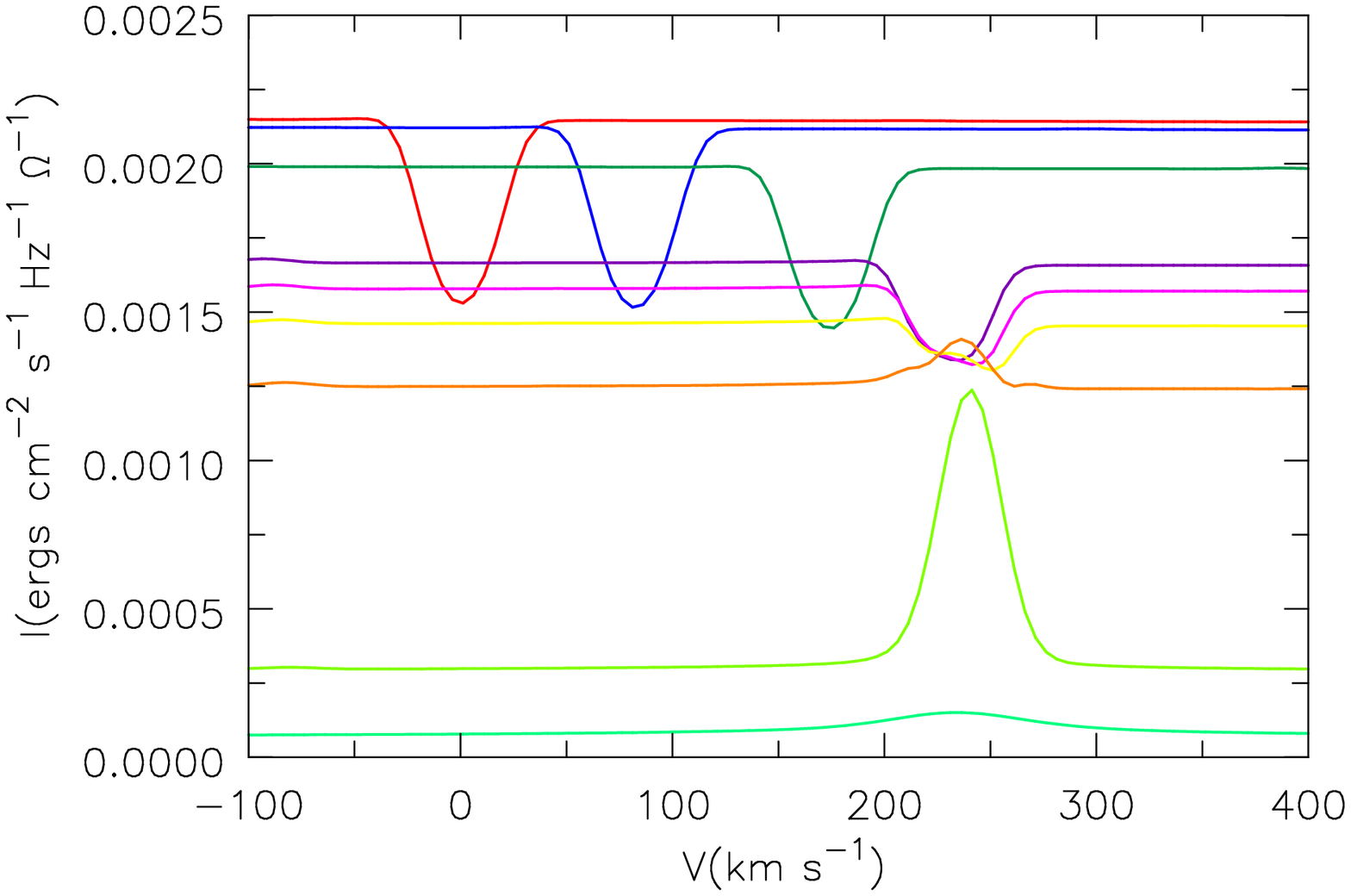}
\hspace{1.5in}\parbox{3.0in}{\caption{As for Fig.~\ref{fig_niv_id} but illustrating \civ\ \lb 5801.
The strength of the emission relative to the continuum is significantly enhanced towards the
limb of the star.}
\label{fig_civ_id} }
\end{minipage}
\begin{minipage}[t]{0.48 \linewidth}
\centering
\includegraphics[scale=0.42, angle=-90]{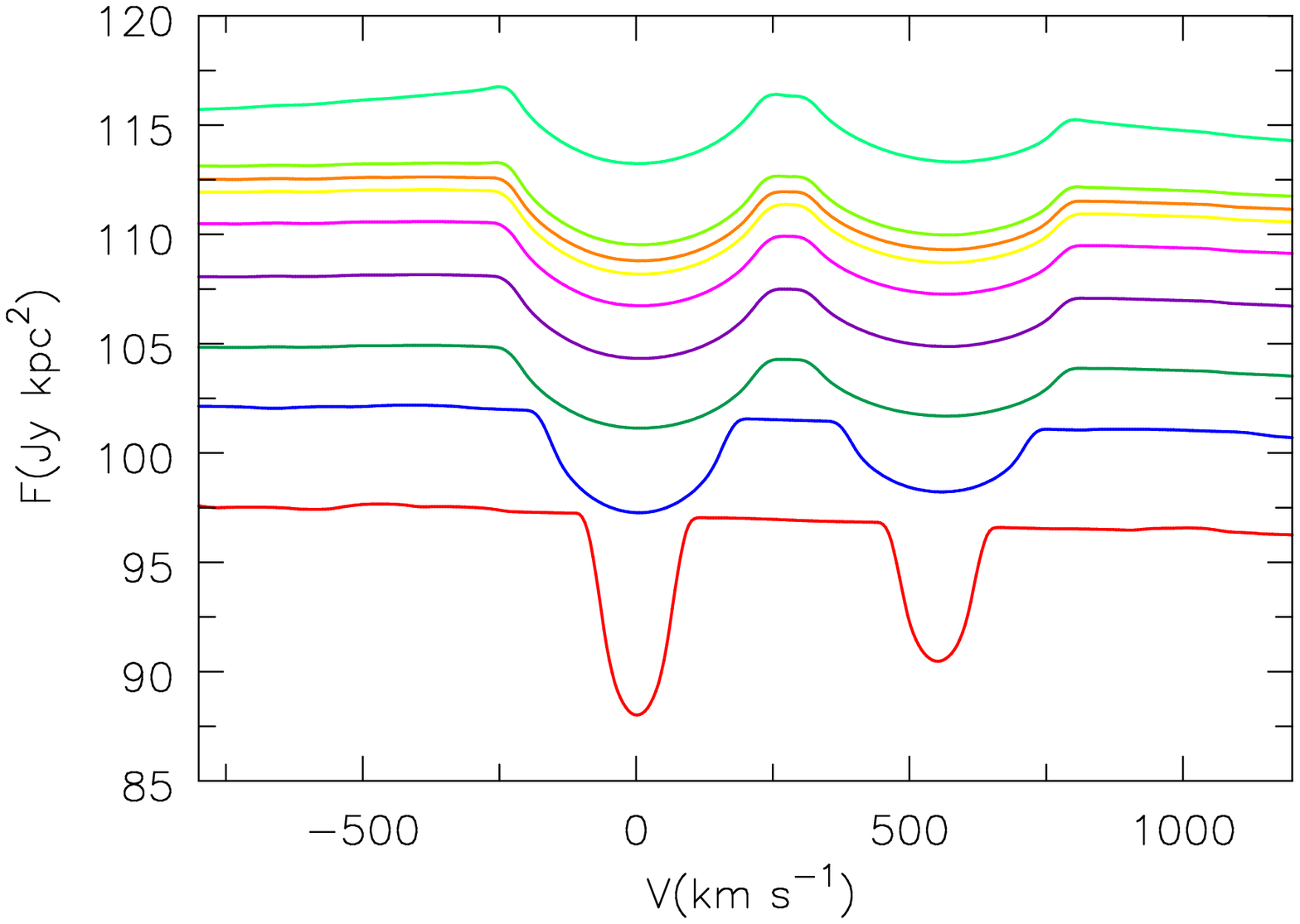}
\hspace{1.5in}\parbox{3.0in}{\caption{As for Fig.~\ref{fig_niv_sp} but illustrating  \civ\ \lb 5801.
At $\sim$500\,\kms we see  the other component (\civ\ \lb 5812) of the doublet. In the top spectrum
an additional broad emission component arising from the wind is evident.}
\label{fig_civ_sp} } 
\end{minipage}
\end{figure*}

To complete our study we illustrate two other lines. The first is \civ\ \lb 5801 which shows an
absorption component  together with extended wind emission. In many O stars this line is very difficult to model, being sensitive to both line blanketing \citep{BHL12_Osg} and model parameters. This line, like \niv\ \lb 4058, is in absorption for core rays (Fig.~\ref{fig_civ_id}), but in emission for rays near the limb. However, in contrast to \niv\ \lb 4058, the line remains in absorption  longer (i.e., to higher impact parameters) before switching to emission. The central part of the line in the full spectrum thus remains in absorption (Fig.~\ref{fig_civ_sp}). Careful study of Fig.~\ref{fig_civ_sp} reveals extended emission line wings arising from the wind. Given the transition from absorption to emission as we move to higher impact parameters, and since what we observed is a weighted sum of this variation, it is not surprising that the line is model sensitive.

From Fig.~\ref{fig_prof} we see that the \civ\ absorption profiles are narrower than
what one would predict based of the rotation rate of the stars. Although not so evident
in the current model, disk profiles for \civ\ \lb 5801 are generally narrower than those
obtained via convolution. To get a better match for the present model would require
stronger \civ\ absorption for core rays, and (possibly) enhanced emission for the
limb rays. Because the absorption is preferentially confined to the core rays, the
full disk absorption profiles would be narrower than for pure absorption lines.

\begin{figure*}
\begin{minipage}[t]{0.48 \linewidth} 
\centering
\includegraphics[scale=0.42, angle=-90]{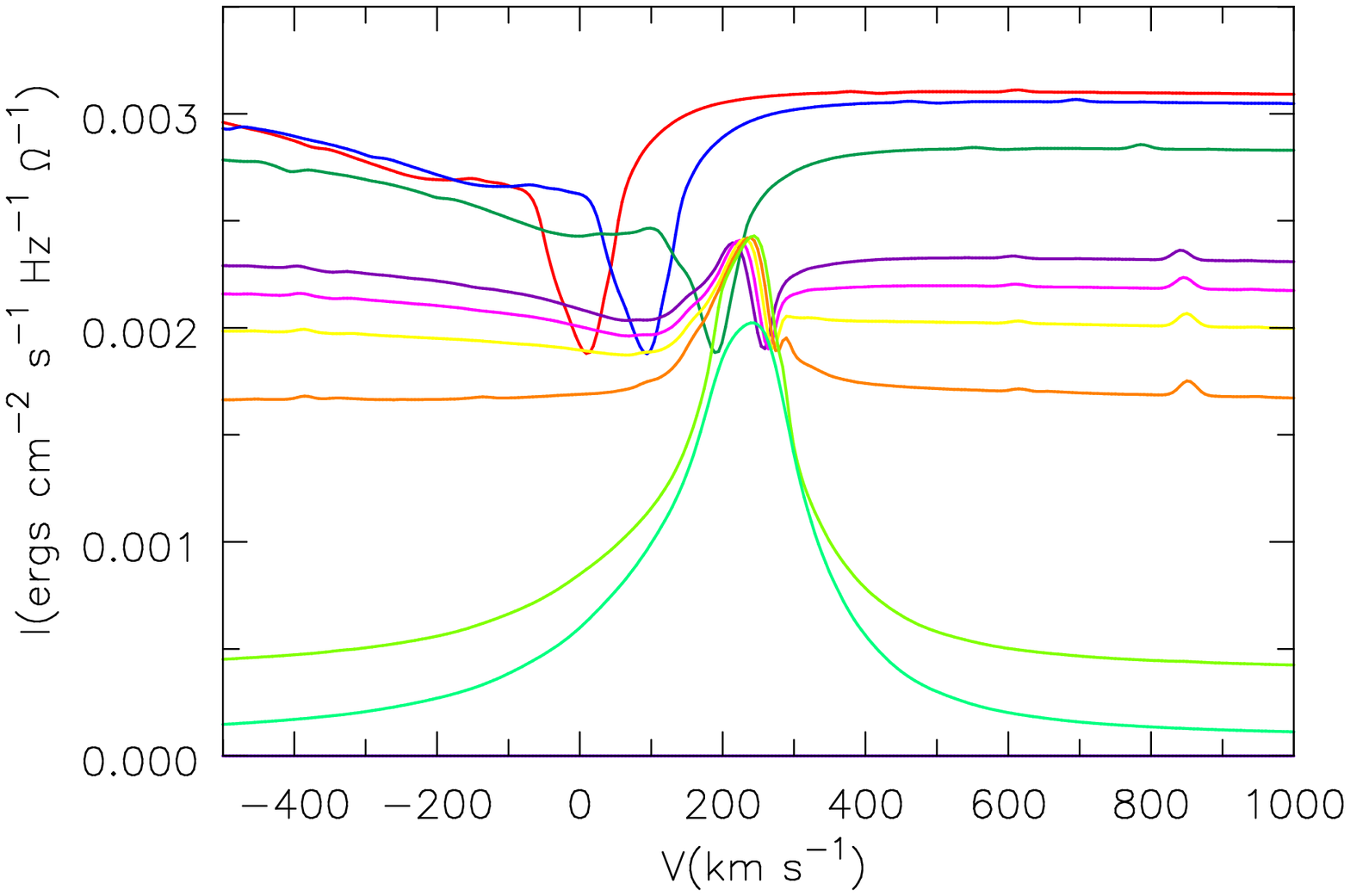}
\hspace{1.5in}\parbox{3.0in}{\caption{As for Fig.~\ref{fig_niv_id} but illustrating \heii\ \lb 4686.
Because of the wind, these profiles exhibit absorption in the blue.}
\label{fig_4686_id} }
\end{minipage}
\begin{minipage}[t]{0.48 \linewidth}
\centering
\includegraphics[scale=0.42, angle=-90]{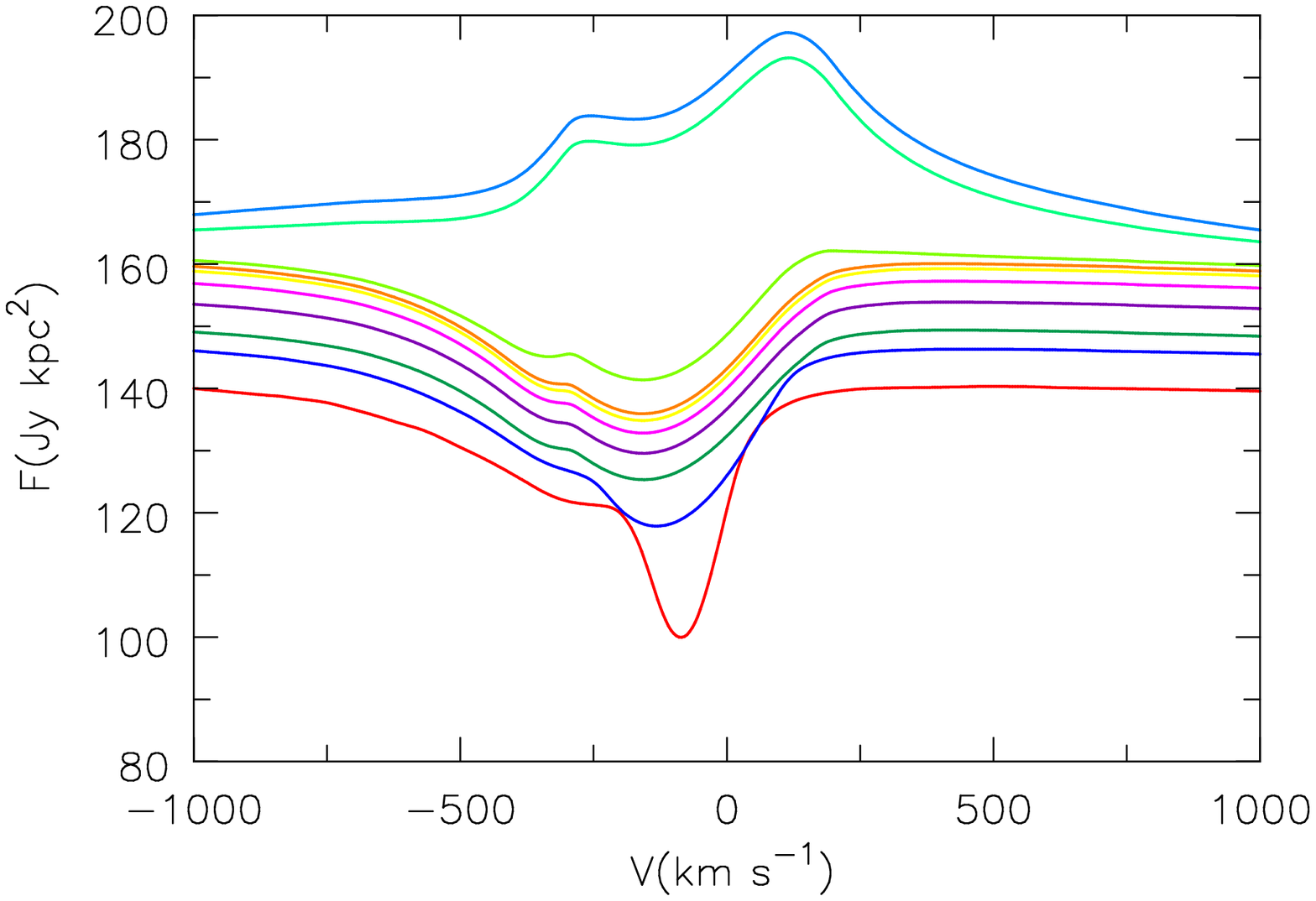}
\hspace{1.5in}\parbox{3.0in}{\caption{As for Fig.~\ref{fig_niv_sp} but illustrating  \heii\ \lb 4686.
The additional curve (second from top) is for $p/\Rstar= 2.513$, and was included because this
line is significantly influenced by wind emission.}
\label{fig_4686_sp} }
\end{minipage}
\end{figure*}

The final line studied is \heii\ \lb 4686. This line exhibits strong emission, but in \zpup\
(and some other stars) it shows a notch on the blue side. As shown in Fig.~\ref{fig_prof},
this notch can be explained by correctly accounting for the effects of rotation. Like \niv\
\lb 4058 and \civ\ \lb 5801 this line also transitions from absorption for low impact parameters
to emission for high impact parameters (Fig.~\ref{fig_4686_id}). However there are important distinctions in its behavior from these two lines. First,  we see extended absorption to the blue, arising from the stellar wind. Second, wind emission is very important and is what drives the line into emission (Fig.~\ref{fig_4686_sp}).

\section{Conclusion}

We have demonstrated that rotation alone can explain the rectangular profiles, sometimes
double-peaked, that are observed  for some optical emission lines in O stars. It is
not necessary to invoke departures of the density structure from spherical symmetry to
explain these profiles.  To properly compute the emission line profiles, it is necessary
to integrate the intensities across the rotating stellar disk -- the classic convolution of the
flux does not work because the line profiles vary substantially across the stellar disc (i.e., with impact parameter). In some cases the line is in absorption for rays striking the center of the star, while it is strongly in emission for rays near the limb. For \niv\ \lb 4058 in \zpup, the observed weak double
peaked profile is actually a result of near perfect cancellation between absorption at low impact parameters, and emission at high impact parameters. This illustrates that the variation in intensity
with impact parameter can provide additional insights into line formation.

From their study of the \niii\ lines in HD 16691, \cite{DRL09_prof} concluded that the lines arise
from close to the star in a large scale corotating structure. While we confirm that rotation is
crucial in explaining the observed line shapes, we find that the \niii\ lines are primely of photospheric origin in these supergiants and that rotation alone can explain their observed shapes.

For most analyses it is adequate to use the normal convolution technique when modeling absorption lines. However, there are subtle differences between disk profiles, and the profiles obtained with convolution, and these could be important when using absorption lines to study line asymmetries and the possible influence of the velocity field at the base of the wind. Following a suggestion by Francisco Najarro, and due to the intense interest with the mass of the most massive stars, we also ran a model for R136a3, which was found by \cite{CSH10_R136} to have an initial mass of $\sim 165\,$\Msun\ and a $v \sin i \sim 200\,$\kms. Profiles computed using a 2D model to take into account rotation showed only small differences from profiles computed using the convolution technique, and are too small to significantly affect conclusions drawn by \cite{CSH10_R136}.

Although already well known, we also highlighted the influence rotation has on  main optical wind diagnostics -- H$\alpha$ and \heii\ \lb 4686. While the correct allowance for rotation does not fundamentally alter  the line strength (and hence, for example, inferred mass-loss rates) it does significantly alter
the line profile. This is of crucial importance -- it means that H$\alpha$ and \heii\ \lb 4686 emission line profiles in rapidly rotating O stars cannot be used to infer variations in clumping with distance and
the importance of velocity-porosity unless rotation is accurately taken into account. Further, rotation will influence the accuracy of parameters, such as $\beta$ (in the classic wind-law $v(r)=(1-r/\Rref)^\beta$). 

The correct treatment of rotation is also important for UV resonance lines formed in the wind. As photospheric lines can influence the resonance line profiles, it is important to take rotation into account. However, a simple convolution overestimates the effect of rotation on the wind lines and can adversely affect estimates of the terminal velocity, and microturbulence within the wind.

Since it is relatively easy to treat the gross influence of rotation accurately, this should become the norm in
future studies of O stars. Issues related to the influence of rotation on the azimuthal density structure, and the dependence of clumping parameters on rotation, will require further studies.

\label{Sec_future}

\section*{acknowledgements}
DJH acknowledges support from NASA ADP Grant: NNG04GC81G, STScI theory
grant HST-AR-11756.01.A and HST-AR-12640.01. STScI is operated by the Association of
Universities for Research in Astronomy, Inc., under NASA contract NAS5-26555. 
J.-C. Bouret thanks the French Agence Nationale de la Recherche (ANR) for financial support.
The authors thank the referee, Joachim Puls,  for his suggestions, and his careful reading of the paper.
The authors would also like to thank Francisco Najarro for comments on a draft of the paper.

\bibliography{bib_rot}

\begin{thebibliography}{21}
\expandafter\ifx\csname natexlab\endcsname\relax\def\natexlab#1{#1}\fi

\bibitem[{{Bjorkman} \& {Cassinelli}(1993)}]{BC93_WCD}
{Bjorkman}, J.~E. \& {Cassinelli}, J.~P. 1993, \apj, 409, 429

\bibitem[{{Bouret} {et~al.}(2012){Bouret}, {Hillier}, {Lanz}, \&
  {Fullerton}}]{BHL12_Osg}
{Bouret}, J.-C., {Hillier}, D.~J., {Lanz}, T., \& {Fullerton}, A.~W. 2012,
  \aap, submitted

\bibitem[{{Bruccato} \& {Mihalas}(1971)}]{BM71_NIII}
{Bruccato}, R.~J. \& {Mihalas}, D. 1971, \mnras, 154, 491

\bibitem[{{Busche} \& {Hillier}(2005)}]{BH05_2D}
{Busche}, J.~R. \& {Hillier}, D.~J. 2005, \aj, 129, 454

\bibitem[{{Crowther} {et~al.}(2010){Crowther}, {Schnurr}, {Hirschi}, {Yusof},
  {Parker}, {Goodwin}, \& {Kassim}}]{CSH10_R136}
{Crowther}, P.~A., {Schnurr}, O., {Hirschi}, R., {Yusof}, N., {Parker}, R.~J.,
  {Goodwin}, S.~P., \& {Kassim}, H.~A. 2010, \mnras, 408, 731

\bibitem[{{De Becker} {et~al.}(2009){De Becker}, {Rauw}, \&
  {Linder}}]{DRL09_prof}
{De Becker}, M., {Rauw}, G., \& {Linder}, N. 2009, \apj, 704, 964

\bibitem[{{Gray}(1992)}]{Gray92_book}
{Gray}, D.~F. 1992, {The observation and analysis of stellar photospheres.}
  (Cambridge University Press)

\bibitem[{{Hillier} {et~al.}(2003){Hillier}, {Lanz}, {Heap}, {Hubeny}, {Smith},
  {Evans}, {Lennon}, \& {Bouret}}]{HLH03_AV83}
{Hillier}, D.~J., {Lanz}, T., {Heap}, S.~R., {Hubeny}, I., {Smith}, L.~J.,
  {Evans}, C.~J., {Lennon}, D.~J., \& {Bouret}, J.-C. 2003, \apj, 588, 1039

\bibitem[{{Howarth} {et~al.}(1995){Howarth}, {Prinja}, \&
  {Massa}}]{HPM95_zeta_pup}
{Howarth}, I.~D., {Prinja}, R.~K., \& {Massa}, D. 1995, \apjl, 452, L65

\bibitem[{{Maeder} \& {Meynet}(2000)}]{MM00_Edd_lim}
{Maeder}, A. \& {Meynet}, G. 2000, \aap, 361, 159

\bibitem[{{Mihalas} {et~al.}(1972){Mihalas}, {Hummer}, \& {Conti}}]{MHC72_NIII}
{Mihalas}, D., {Hummer}, D.~G., \& {Conti}, P.~S. 1972, \apjl, 175, L99

\bibitem[{{Najarro} {et~al.}(2011){Najarro}, {Hanson}, \& {Puls}}]{NHP11_Lband}
{Najarro}, F., {Hanson}, M.~M., \& {Puls}, J. 2011, \aap, 535, A32

\bibitem[{{Owocki} {et~al.}(1998){Owocki}, {Cranmer}, \&
  {Gayley}}]{OCG98_WCD_inhib}
{Owocki}, S.~P., {Cranmer}, S.~R., \& {Gayley}, K.~G. 1998, \apss, 260, 149

\bibitem[{{Petrenz} \& {Puls}(1996)}]{PP96_2D}
{Petrenz}, P. \& {Puls}, J. 1996, \aap, 312, 195

\bibitem[{{Petrenz} \& {Puls}(2000)}]{PP00_2D}
---. 2000, \aap, 358, 956

\bibitem[{{Puls} {et~al.}(2006){Puls}, {Markova}, {Scuderi}, {Stanghellini},
  {Taranova}, {Burnley}, \& {Howarth}}]{PMS06_clump}
{Puls}, J., {Markova}, N., {Scuderi}, S., {Stanghellini}, C., {Taranova},
  O.~G., {Burnley}, A.~W., \& {Howarth}, I.~D. 2006, \aap, 454, 625

\bibitem[{{Rivero Gonz{\'a}lez} {et~al.}(2011){Rivero Gonz{\'a}lez}, {Puls}, \&
  {Najarro}}]{RPN11_NIII}
{Rivero Gonz{\'a}lez}, J.~G., {Puls}, J., \& {Najarro}, F. 2011, \aap, 536, A58

\bibitem[{{Schilbach} \& {R{\"o}ser}(2008)}]{SR08_Ostars}
{Schilbach}, E. \& {R{\"o}ser}, S. 2008, \aap, 489, 105

\bibitem[{{Sota} {et~al.}(2011){Sota}, {Ma{\'{\i}}z Apell{\'a}niz}, {Walborn},
  {Alfaro}, {Barb{\'a}}, {Morrell}, {Gamen}, \& {Arias}}]{SMW11_class}
{Sota}, A., {Ma{\'{\i}}z Apell{\'a}niz}, J., {Walborn}, N.~R., {Alfaro}, E.~J.,
  {Barb{\'a}}, R.~H., {Morrell}, N.~I., {Gamen}, R.~C., \& {Arias}, J.~I. 2011,
  \apjs, 193, 24

\bibitem[{{Sundqvist} {et~al.}(2011){Sundqvist}, {Puls}, {Feldmeier}, \&
  {Owocki}}]{SPF11_clumps}
{Sundqvist}, J.~O., {Puls}, J., {Feldmeier}, A., \& {Owocki}, S.~P. 2011, \aap,
  528, A64

\bibitem[{{Walborn}(1971)}]{Wal71_Of}
{Walborn}, N.~R. 1971, \apjs, 23, 257

\end{thebibliography}

\label{lastpage}
\end{document}